\documentclass[aps,twocolumn,showpacs,preprintnumbers,amsmath,amssymb]{revtex4}
 \usepackage{graphicx}
 \usepackage[pdfstartview=FitH, colorlinks=true, linkcolor=blue, citecolor=red]{hyperref}

 \def\T{\textstyle}
 \def\l{\left}
 \def\r{\right}
 \def\nf{n_{\!f}}	

 \def\be{\begin{equation}}
 \def\ee{\end{equation}}
 \def\bea{\begin{eqnarray}}
 \def\eea{\end{eqnarray}}
 \def\bean{\begin{eqnarray*}}
 \def\eean{\end{eqnarray*}}

 \def\gsim{\mathrel{\rlap{\lower0.2em\hbox{$\sim$}}\raise0.2em\hbox{$>$}}}
 \def\ksim{\mathrel{\rlap{\lower0.2em\hbox{$\sim$}}\raise0.2em\hbox{$<$}}}
\begin{document}

\title{Running coupling and screening in the (s)QGP}

\author{A.~Peshier}

\affiliation{
 Institut f\"{u}r Theoretische Physik,
 Universit\"{a}t Giessen, 35392 Giessen, Germany}


\begin{abstract} \noindent
Emphasizing the importance of renormalization in the context of thermal field theory in general, it is pointed out that the Debye mass in the hot quark gluon plasma is determined by the coupling at the scale $m_D$, not $T$ as commonly presumed. The mended result agrees quantitatively with lattice QCD calculations in the strong coupling regime almost down to $T_c$.
\end{abstract}

\maketitle

\section{Introduction \label{sec: intro}}
It is common knowledge that the presence of a medium does not introduce new UV divergencies in quantum field theoretical calculations, basically because the momenta of the medium fluctuations are constrained by the temperature $T$ and/or the chemical potential $\mu$.
At times this fact is used to comfort a somewhat pragmatic strategy in practical calculations, namely discussing only the matter contribution to the quantity under consideration, deferring the renormalization (of the vacuum part) as standard `textbook physics'.
Such a procedure, however, should be applied with the appropriate care in cases where new momentum scales occur, as in the context of statistical field theory. A related aspect thereof -- which is crucial for phenomenology -- is the arising ambiguity in the coupling, whose momentum scale then is not fixed by the calculation, but has to be chosen somewhat {\em ad hoc}.

A widely used expression obtained from such a pragmatic prescription is
\be
  \tilde m_D^2
  = 
  \l( 1+\T\frac16 \nf \r) 4\pi\alpha(Q_T^2)\, T^2 \quad
  \label{eq: mD pre}
\ee
for the Debye mass in a hot quark gluon plasma (QGP) with $\nf$ light quark flavors.
Describing the screening of (chromo-) electric fields, the Debye mass is a key parameter for a variety of phenomenologically relevant quantities. 
For quantitative estimates, the coupling $\alpha$ has prevalently been fixed at a `typical' thermal scale of the order of the (bosonic) Matsubara frequency, $Q_T = c \cdot 2\pi T$, with $c \sim 1$.
Recently, the expression (\ref{eq: mD pre}) has also been utilized as a parameterization of lattice QCD (lQCD) results \cite{Kaczmarek:2004gv, Nakamura:2003pu}. It there turned out that the Debye mass in the non-perturbative regime near the (phase-) transition temperature $T_c$ can be reasonably described by a re-scaling {\em Ansatz} $\kappa \tilde{m}_D$, see below.

A reckoned explanation of the observed enhancement, $\kappa \approx 1.5$, is the next-to-leading order correction \cite{Rebhan:1993az} to the result (\ref{eq: mD pre}) which, under reasonable assumptions for the magnetic mass, is positive.
As an aside it is noted that such an interpretation is quite non-trivial. Naively, one could expect that if $m_D/T \sim (4\pi\alpha)^{1/2} = g$ changes notably in a certain temperature range and the next-to-leading order correction is of a similar size as the leading term, one cannot parameterize the modification by a {\em constant} factor.
However, in QCD the correction is logarithmically enhanced due to an IR sensitivity of the Debye mass, implying
$
	\kappa^{nlo}
	=
	1 + Ag\l( \ln g^{-1} + c \r) \, ,
$
where $A = 3/\l( 4\pi(1+\frac16\nf)^{1/2} \r)$ \cite{Rebhan:1993az}. Thus, from $\Delta\kappa^{nlo} = (\partial\kappa^{nlo}\!/\partial g) \Delta g = (\kappa^{nlo}-1-Ag) \Delta g/g$, a sizable change in $g$ can indeed be reconciled with an almost constant enhancement factor if $\kappa^{nlo}-1-Ag \approx 0$. 
Although this condition is fulfilled in the cases studied below, it turned out that the next-to-leading order correction can only partly account for the observed excess \cite{Nakamura:2003pu}.

Pursued here is a more elementary interpretation of the large Debye mass near $T_c$, namely by questioning the (temperature dependence of the) coupling in (\ref{eq: mD pre}). 
In order to first define rigorously the notion of the running coupling for the following considerations, some basic facts about the coupling renormalization (at $T = 0$) are recollected in Sec.\ \ref{sec: RC}.
Then in Sec.~\ref{sec: mD}, the pole of the propagator at $T>0$ is expressed in terms of the renormalized coupling. By specifying the general result for the case of the Debye mass it is shown that the presupposed scale $Q_T$ in the coupling in Eq.~(\ref{eq: mD pre}) is not justified. 
In some places the notation is kept schematic, both for simplicity and to indicate that the considerations are, in principle, not restricted to QCD (albeit consequences for weakly coupled systems are only marginal). Although the framework is entirely perturbative, it will be instructive -- with regard to the so-called strongly coupled QGP (sQGP) -- to confront the results with lQCD calculations. The paper concludes with a summary and some implications.

\section{Running coupling \label{sec: RC}}
\begin{figure}[ht]
 \hskip-4mm \includegraphics[width=7cm]{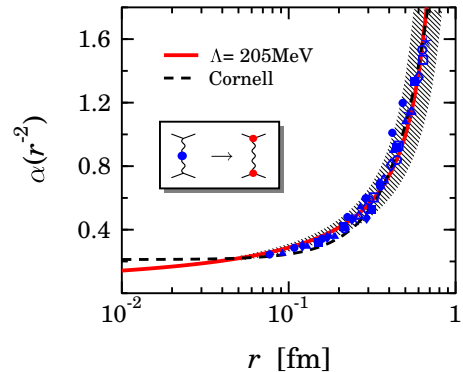}
 \caption{Running coupling at $T=0$, for $\nf = 2$. 
 	 Compared are the lQCD results \cite{Kaczmarek:2004gv} (given here
 	 without error bars) to Eq.~(\ref{eq: e2 run}) with 
 	 $t \rightarrow r^{-2}$ (full line); the hatched band depicts a
 	 variation $\Delta\Lambda = 40\,$MeV.
	 The dashed (black) line represents the result \cite{Kaczmarek:2004gv}
	 from the Cornell parameterization of $V(r)$, which for 
	 $r \gsim 0.3\,$fm almost coincides with the full (red) line.
 \label{fig: alpha(t)}}
\end{figure}

To relate the bare coupling $\alpha$ to an experimental observable,
consider a scattering process in a kinematic regime dominated by the $t$-channel contribution. 
Considering for simplicity a massless theory, the relevant part of the matrix element is $\alpha/\l( P^2-\Pi_{\rm vac}(P^2) \r)$.
To 1-loop order and in dimensional regularization, the boson selfenergy has the generic form $\Pi_{\rm vac}(P^2) = \alpha P^2 \l( \epsilon^{-1} - \ln(-P^2/\mu^2) \r)$.
Then, for a specific $P^2 = t_e$, the matrix element reads
\be
 \frac1{t_e}\, \frac\alpha{1-\alpha\l( \epsilon^{-1}-\ln(-t_e/\mu^2) \r)}
 \equiv
 \frac{\alpha(t_e)}{t_e} \, .
 \label{eq: M}
\ee
The right hand side can be considered as experimental input, introducing the coupling $\alpha(t_e)$ at the scale $t_e$, which is related to the (infinite) bare coupling by
\be
 \alpha^{-1}(t_e) = \alpha^{-1} - \epsilon^{-1} + \ln(-t_e/\mu^2) \, .
 \label{eq: e2 vs e2(t)}
\ee
An equivalent relation holds for the coupling $\alpha(t)$ at an arbitrary scale $t$, consequently
\[
  \alpha^{-1}(t) = \alpha^{-1}(t_e) + 4\pi\beta_0\, \ln(t/t_e) \, .
\]
Here the previously omitted prefactor has been re-introduced, which determines the asymptotic behavior of the specific theory under consideration. For QCD with $N_c = 3$ colors and $\nf$ light quark flavors, the leading order coefficient of the $\beta$-function is $\beta_0 = (\frac{11}3\, N_c - \frac23\, \nf)/(4\pi)^2 > 0$.
The momentum dependence of the `running' coupling is fully specified by its value at the scale $t_e$, or in a common alternative way of writing,
\be
  \alpha(t) = \frac1{4\pi\beta_0}\, \frac1{\ln(|t|/\Lambda^2)} \, ,
  \label{eq: e2 run}
\ee
by the QCD parameter $\Lambda$.
It is also noted that the matrix element (\ref{eq: M}) can now be expressed simply as
\[
 \frac\alpha{P^2-\Pi_{\rm vac}(P^2)}
 =
 \frac{\alpha(P^2)}{P^2} \, ,
\]
with an obvious diagrammatic interpretation as depicted in the insert of Fig.~\ref{fig: alpha(t)}.

Before turning to the case of finite temperature, it is instructive to challenge the 1-loop expression for the running coupling by a comparison to non-perturbative lQCD results.
To that end, Eq.~(\ref{eq: e2 run}) is applied in the position space by replacing $|t| \rightarrow r^{-2}$. On the lattice, the coupling can be obtained from the heavy quark potential $V(r)$ or more conveniently, as advocated in \cite{Kaczmarek:2004gv} and references therein, from the corresponding force,
\[
	\alpha_{qq}(r) = \frac34\, r^2 \frac{dV(r)}{dr} \, .
\]
The comparison in Fig.~\ref{fig: alpha(t)} shows an agreement which is at least as good as that obtained in \cite{Kaczmarek:2004gv} from the Cornell parameterization of the potential, $V^{\rm cp}(r) = -\frac43\, a/r + \sigma r$, with a fitted constant $a$ and the string tension $\sqrt\sigma = 420\,$MeV.
Worth to note is that the adjusted value $\Lambda \approx 205\,$MeV for the running coupling (\ref{eq: e2 run}) is verily in the expected ball park. This demonstrates that the pronounced increase of $\alpha_{qq}(r)$ is not an unambiguous indication of non-perturbative effects, as $\alpha^{\rm cp}(r) \rightarrow \frac34\sigma r^2$, but can be utterly accommodated within the asymptotic form, at least at intermediate distances.
This is a motivation that phenomenological considerations based on the 1-loop running coupling may be viable even for moderate interaction strength, say $\alpha \ksim 0.6$ (corresponding to $r \ksim 0.3\,$fm) as relevant in the following.

\section{Debye mass \label{sec: mD}}
The Debye screening mass can be calculated from the position of a pole of the dressed propagator in the presence of a thermal medium. Here the selfenergy receives an additional finite contribution,
\be
  \Pi
  =
  \alpha\l[
   P^2 \l( \epsilon^{-1} - \ln(-P^2/\mu^2) \r) + f(p_0,p)
  \r] ,
  \label{eq: Pi T}
\ee
which in general depends separately on $p_0$ and $p$.
The pole is determined by $p_0^2-p^2-\Pi(p_0,p) = 0$, which can also be written as
\[
  \alpha^{-1}
  =
  \epsilon^{-1} - \ln(-P^2/\mu^2) + f(p_0,p)/P^2\, .
\]
Recognizing in this expression the running coupling $\alpha(P^2)$, cf.\
Eq.~(\ref{eq: e2 vs e2(t)}), this relation is cast in the intuitive form
\be
  P^2 = \alpha(P^2)\, f(p_0,p) \, .
  \label{eq: pole}
\ee
While simple, this equation is remarkable since it is manifestly finite. The pole is determined only by the matter part of the selfenergy, whereas the (divergent) vacuum contribution is `absorbed' completely in the renormalized coupling.
The fact that the pole depends only on the measurable parameter $\Lambda$ (while the auxiliary scale $\mu$ has dropped out) is a strong formal indication that the approximation (\ref{eq: pole}) has a direct physical meaning \footnote{
	Similarly, for gauge theories, the propagator in general depends on
	the gauge but -- as a requirement for a reasonable approximation --
	the pole should be invariant \cite{Kobes:1990xf}.}.

As a special case of Eq.~(\ref{eq: pole}), the Debye screening mass is defined as the pole position of the longitudinal boson propagator $D_{00}$ for $p_0 = 0$, at space-like momentum $-P^2 = m_D^2$.
Then it is first obvious that screening is an effect solely due to the thermal fluctuations. The vacuum contributions of the selfenergy (\ref{eq: Pi T}) do not have to be `omitted', e.\,g., by misleadingly arguing that for large $T$ they are exceeded (after renormalization) by the matter part, $f \sim T^2 \gg p^2$. On the contrary, they are actually necessary to render finite the resulting equation.
Not surprising, in fact, is that the relevant scale in the running coupling is given by the Debye mass itself, viz
\be
  m_D^2 = \T\frac13N_c \l( 1+\frac16 \nf \r) 4\pi\alpha(m_D^2)\, T^2 \, .
  \label{eq: mD def}
\ee
Here the same perturbative result for $f(0, p)$ as in Eq.~(\ref{eq: mD pre}) has been used, which for small $p$ is momentum independent and explicitly gauge invariant \cite{LeBellac}.
Thus already to leading order (i.\,e.\ without using a resummed selfenergy calculated from dressed propagators), due to renormalization the Debye mass is determined by an implicit equation. 
With the coupling (\ref{eq: e2 run}), the self-consistent solution can be given in terms of Lambert's $W$-function \footnote{
	The Lambert $W$-function, $y=W(x)$, is usually defined as the solution
	of $y \exp(y) = x$, see e.\,g., 
	{\tt mathworld.wolfram.com/LambertW-function.html.}},
\be
	m_D^2 = \frac{bT^2}{W(bT^2/\Lambda^2)} \, ,
  \label{eq: mD}
\ee
where $b = \frac{N_c}3 (1+\frac16 \nf)/\beta_0$.
Even without having to refer to the tabulated properties of the $W$-function, two basic features of the solution can be easily inferred directly from Eq.~(\ref{eq: mD def}). First, $m_D/T$ is a monotonically decreasing function of $T/\Lambda$ and second, the self-consistent solution is larger than the expression (\ref{eq: mD pre}) if $m_D < Q_T$.

\begin{figure}[ht]
 \hskip-4mm \includegraphics[width=7cm]{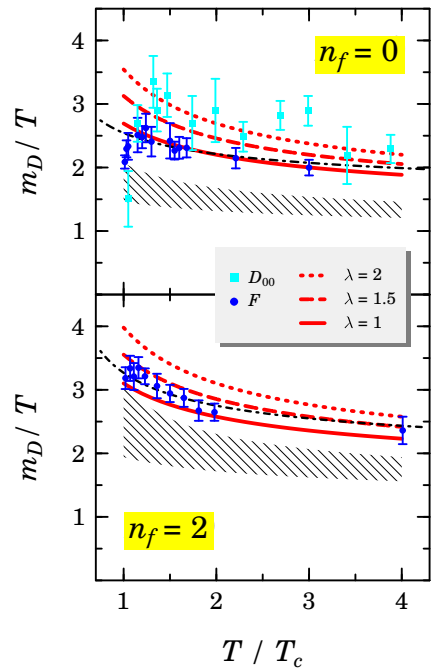}
 \caption{The QCD Debye mass for $\nf=0$ and $\nf=2$.
 		The lQCD results labeled $F$ and $D_{00}$ are from
 		\cite{Kaczmarek:2004gv}	and \cite{Nakamura:2003pu}, respectively.
  	The hatched bands depict Eq.~(\ref{eq: mD pre}) with the 
  	2-loop coupling fixed at $Q_T = c \cdot 2\pi T$ with 
  	$c \in [\frac12,2]$ and values of $\Lambda/T_c$ as specified 
  	in the text.
  	The dash-dotted (black) lines represent the re-scaling fits
  	\cite{Kaczmarek:2004gv} with the {\em Ansatz}	$\kappa \tilde m_D$;
  	here $c=1$.		
  	The thicker (red) lines show the self-consistent result 
  	(\ref{eq: mD}) for some representative choices of the parameter 
  	$\lambda = \Lambda/T_c$.
 	\label{fig: mD}}
\end{figure}

\begin{figure}[ht]
 \hskip-4mm \includegraphics[width=7cm]{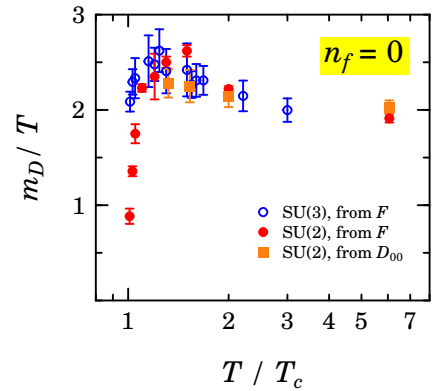}
 \caption{Comparison of quenched lQCD results: SU(3)
   \cite{Kaczmarek:2004gv} vs.\ SU(2) \cite{Digal:2003jc}.
 	 The deviations very near $T_c$ are not unexpected due to the different
 	 order of the phase transition.
 	\label{fig: mD SU(N)}}
\end{figure}

Although also Eq.~(\ref{eq: mD}) is justified {\em a priori} only for $T \gg \Lambda$, it is remunerating to compare it to lQCD results. 
In Monte Carlo calculations, the Debye mass can be obtained by two methods: from analyzing the large-distance behavior of (i) the heavy quark free energy $F$ or (ii) the static electric propagator $D_{00}$.
For quenched QCD as well as for pure SU(2), the Debye mass has been calculated within both approaches. Even though for SU(3) the results  are not fully consistent, see Fig.\ \ref{fig: mD}, they are clearly underestimated, by the aforementioned factor $\kappa \approx 1.5$, by the expression (\ref{eq: mD pre}) supplemented with the 2-loop running coupling \footnote{
	It seems more appropriate to use the 1-loop running coupling in
	connection with the 1-loop formula for $m_D$, what indeed slightly
	reduces the observed discrepancy.}.
The new formula (\ref{eq: mD}), on the other hand, can describe the lQCD results almost down to $T_c$ by appropriate choices of the dimensionless parameter $\lambda = \Lambda/T_c$.
I will focus on the data sets \cite{Kaczmarek:2004gv} since they, in the quenched limit, match nicely with the findings in SU(2) \cite{Digal:2003jc}, cf.\ Fig.~\ref{fig: mD SU(N)}. 
In this computationally less demanding case the lattice results derived from the free energy and from the propagator are indeed compatible.
On the other hand, for $\nf = 0$ the Debye mass (\ref{eq: mD}) is also $N_c$-independent.

Adjusting the scale parameter in Eq.~(\ref{eq: mD}) to the QCD ($N_c = 3$) lattice results (excluding a narrow interval above $T_c$) yields $\lambda_{(\nf=0)} \approx 1$ for the quenched case and $\lambda_{(\nf=2)} \approx 1.3$ for two light flavors. 
One might suspect that the lattice results could be accommodated within this strictly perturbative approach only at the expense of the interpretability of the parameter. I thus emphasize that the fitted $\lambda$ are in a remarkable agreement with the known ratio of the QCD scale to the transition temperature \cite{Kaczmarek:2004gv}, $\Lambda/T_c = 1/1.14(4)$ and $\Lambda/T_c = 1/0.77(21)$ for $\nf = 0$ and $\nf = 2$, respectively.
To round up the emerging consistent picture it is underlined that $\Lambda \approx 205\,$MeV, as adjusted in Sec.~\ref{sec: RC} for the $\nf = 2$ QCD potential at $T = 0$, is (together with the commonly expected value of $T_c \approx 160$MeV) also compatible with $\lambda_{(\nf=2)}$.

Due to the monotonic behavior of $m_D/T$ alluded to before, Eq.~(\ref{eq: mD}) cannot reproduce the characteristic decrease of the Debye mass very close to $T_c$. A possible interpretation has been given as a Boltzmann suppression of the time-like excitations (quasiparticles), which become heavy near $T_c$ \cite{Peshier:1995ty}.
Within the present approach, it could alternatively indicate a non-monotonous $\alpha(Q)$.

\begin{figure}[ht]
 \hskip-4mm \includegraphics[width=7cm]{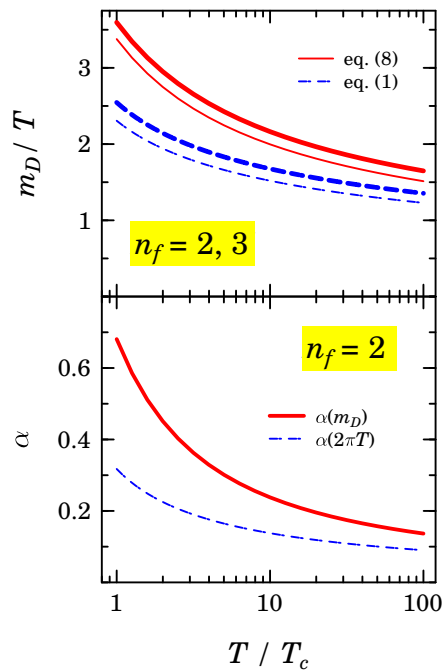}
 \caption{Top: Large-$T$ behavior of the Debye mass for $\nf = 2$ (thin
  lines) and $\nf = 3$ (assuming $\Lambda_{(\nf=3)} =
  \Lambda_{(\nf=2)}$; bold lines). 
 	Bottom: The adjusted 1-loop coupling (\ref{eq: e2 run}) at the scale
 	$m_D$	vs.\ the 2-loop coupling at $2\pi T$.
 	\label{fig: large T}}
\end{figure}

For large temperatures, the deviations of the results from Eqs.\ (\ref{eq: mD pre}) and (\ref{eq: mD}) decrease logarithmically, reaching the order of 10\% only at $10^2T_c$, cf.\ Fig.~\ref{fig: large T}.
From the slopes of the curves it can be expected that the re-scaling {\em Ansatz} $\kappa \tilde{m}_D$ will overestimate the correct Debye mass at high temperatures (a hint thereof may in fact be spotted already in Fig.~\ref{fig: mD}).
Shown also in Fig.~\ref{fig: large T} is the prediction for $\nf = 3$ massless flavors, which should constrain the Debye mass in the physical case ($\nf = 2+1$).

\section{Conclusions}
In a wider sense, it was demonstrated that there is no coupling in a (thermal) medium {\em per se} -- it always is to be related to a specific quantity/scale. The correct scale, which should be an outcome of the calculation rather than `chosen', is crucial in particular for larger coupling.
The relevant scale for the Debye mass,
\[
	m_D^2 \;
	\sim \;\;
	\raisebox{-4mm}{\includegraphics[width=17mm]{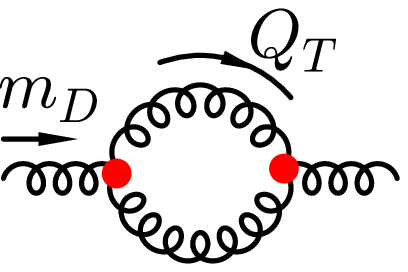}}\;
	+ \ldots
	\;\sim\;
	\alpha(m_D^2) T^2 \, ,
\]
is $m_D$ itself, not the typical loop momentum $Q_T \sim 2\pi T$ as often presumed.
It's the interplay between vacuum and thermal fluctuations (accounted for by renormalization) that leads to the implicit Eq.~(\ref{eq: mD def}), which is manifestly finite and gauge and renormalization group invariant. The soft(er) scale in the coupling, $m_D \sim \sqrt\alpha T \ll T$ for $\alpha \ll 1$, yields stronger screening than previously estimated from (\ref{eq: mD pre}). 
Remarkably, both the novel result (\ref{eq: mD}) and the underlying running coupling (at $T=0$) are consistent with corresponding lQCD results.
One might speculate whether Eq.~(\ref{eq: mD}) is especially suited for extrapolation by its formal amenities -- higher order terms could basically cancel each other if perturbative expansions have similar properties as asymptotic series \cite{Peshier:1998ei}.

Exemplifying the renormalization at $T>0$ is (at the considered level) particularly straightforward for the Debye mass. Yet this quantity is a regulator for many observables with an IR sensitivity and thus central in statistical field theory. Accordingly, there are numerous phenomenological implications for heavy ion physics.
Mentioned here explicitly with regard to the establishing notion of a sQGP is that the electric sector of QCD indeed seems to be quite strongly coupled, $\alpha_E \sim \alpha(m_D) \approx 0.6$ near $T_c$, see Fig.~\ref{fig: large T}.
In the magnetic sector, on the other hand, the effective coupling apparently remains much smaller, $\alpha_M \approx 0.3$ near $T_c$ \cite{Laine:2005ai}.
This indicates a rather nuanced picture of the sQGP, see e.\,g.\ \cite{Oswald:2005vr}.
\\[3mm]
{\bf Acknowledgment:} I thank W.~Cassing, S.~Leupold, R.~Pisarski and especially S.~Peign\'e for fertile discussions, and J.~Noronha for pointing out the relation of Eq. (\ref{eq: mD def}) to the $W$-function.
This work was supported by BMBF.

\end{document}